%
%

\input harvmac
\noblackbox


\font\ticp=cmcsc10 
 
\def\Title#1#2{\rightline{#1}\ifx\answ\bigans\nopagenumbers\pageno0\vskip1in
\else\pageno1\vskip.8in\fi \centerline{\titlefont #2}\vskip .5in}

\font\ticp=cmcsc10
\font\ttsmall=cmtt10 at 8pt

\input epsf
\ifx\epsfbox\UnDeFiNeD\message{(NO epsf.tex, FIGURES WILL BE
IGNORED)}
\def\figin#1{\vskip2in}
\else\message{(FIGURES WILL BE INCLUDED)}\def\figin#1{#1}\fi
\def\ifig#1#2#3{\xdef#1{Fig.\the\figno}
\goodbreak\topinsert\figin{\centerline{#3}}%
\smallskip\centerline{\vbox{\baselineskip12pt
\advance\hsize by -1truein\noindent{\bf Fig.~\the\figno:} #2}}
\bigskip\endinsert\global\advance\figno by1}


%
%

\def\fun#1#2{\lower3.6pt\vbox{\baselineskip0pt\lineskip.9pt
  \ialign{$\mathsurround=0pt#1\hfil##\hfil$\crcr#2\crcr\sim\crcr}}}
\relax

\def\[{\left [}
\def\]{\right ]}
\def\({\left (}
\def\){\right )}

\def\ie{{\it i.e.}}

\def\p{\partial}

\def\b{\beta}
\def\a{\alpha}

\def\CO{{\cal O}}

\def\CN{{\cal N}}

\def\S{{\bf S}}

\def\A2S2{AdS$_2 \times  \S^2$}

\def\ie{{\it i.e.}}
\def\Up{\Upsilon}

\def\bUp{{\bar \Upsilon}}
\def\bY{{\bar Y}}
\def\bF{{\bar F}}
\def\kt{$ K3 \times T^2$}

\lref\sen{
A.~Sen,
{\it Extremal black holes and elementary string states},
Mod.\ Phys.\ Lett.\ A {\bf 10}, 2081 (1995)
[arXiv:hep-th/9504147].
}

\lref\atish{
A.~Dabholkar,
{\it Exact counting of black hole microstates},
[arXiv:hep-th/0409148].
}

\lref\fks{
S.~Ferrara, R.~Kallosh and A.~Strominger,
{\it N=2 extremal black holes},
Phys.\ Rev.\ D {\bf 52}, 5412 (1995)
[arXiv:hep-th/9508072].

A.~Strominger,
``Macroscopic Entropy of $N=2$ Extremal Black Holes,''
Phys.\ Lett.\ B {\bf 383}, 39 (1996)
[arXiv:hep-th/9602111].

S.~Ferrara and R.~Kallosh,
{\it Supersymmetry and Attractors},
Phys.\ Rev.\ D {\bf 54}, 1514 (1996)
[arXiv:hep-th/9602136].

S.~Ferrara and R.~Kallosh,
{\it Universality of Supersymmetric Attractors},
Phys.\ Rev.\ D {\bf 54}, 1525 (1996)
[arXiv:hep-th/9603090].

}

\lref\cdm{
K.~Behrndt, G.~Lopes Cardoso, B.~de Wit, D.~Lust, T.~Mohaupt and W.~A.~Sabra,
{\it Higher-order black-hole solutions in N = 2 supergravity and Calabi-Yau
string backgrounds},
Phys.\ Lett.\ B {\bf 429}, 289 (1998)
[arXiv:hep-th/9801081].

G.~Lopes Cardoso, B.~de Wit and T.~Mohaupt,
{\it Corrections to macroscopic supersymmetric black-hole entropy},
Phys.\ Lett.\ B {\bf 451}, 309 (1999)
[arXiv:hep-th/9812082].

G.~Lopes Cardoso, B.~de Wit and T.~Mohaupt,
{\it Deviations from the area law for supersymmetric black holes},
Fortsch.\ Phys.\  {\bf 48}, 49 (2000)
[arXiv:hep-th/9904005].

G.~Lopes Cardoso, B.~de Wit and T.~Mohaupt,
{\it Macroscopic entropy formulae and non-holomorphic corrections for
supersymmetric black holes},
Nucl.\ Phys.\ B {\bf 567}, 87 (2000)
[arXiv:hep-th/9906094].

G.~Lopes Cardoso, B.~de Wit and T.~Mohaupt,
{\it Area law corrections from state counting and supergravity},
Class.\ Quant.\ Grav.\  {\bf 17}, 1007 (2000)
[arXiv:hep-th/9910179].
}

\lref\cdkm{
G.~L.~Cardoso, B.~de Wit, J.~Kappeli and T.~Mohaupt,
{\it Supersymmetric black hole solutions with R**2 interactions},
[arXiv:hep-th/0003157].

G.~Lopes Cardoso, B.~de Wit, J.~Kappeli and T.~Mohaupt,
{\it Stationary BPS solutions in N = 2 supergravity with R**2 interactions},
JHEP {\bf 0012}, 019 (2000)
[arXiv:hep-th/0009234].

G.~L.~Cardoso, B.~de Wit, J.~Kappeli and T.~Mohaupt,
{\it Examples of stationary BPS solutions in N = 2 supergravity theories  with
R**2-interactions},
Fortsch.\ Phys.\  {\bf 49}, 557 (2001)
[arXiv:hep-th/0012232].
}

\lref\mohaupt{
T.~Mohaupt,
``Black hole entropy, special geometry and strings,''
Fortsch.\ Phys.\  {\bf 49}, 3 (2001)
[arXiv:hep-th/0007195].
}

\lref\wald{
R.~M.~Wald,
{\it Black hole entropy in the Noether charge},
Phys.\ Rev.\ D {\bf 48}, 3427 (1993)
[arXiv:gr-qc/9307038].
}

\lref\osv{
H.~Ooguri, A.~Strominger and C.~Vafa,
{\it Black hole attractors and the topological string},
[arXiv:hep-th/0405146].
}

\lref\vafacy{
C.~Vafa,
{\it Black holes and Calabi-Yau threefolds},
Adv.\ Theor.\ Math.\ Phys.\  {\bf 2}, 207 (1998)
[arXiv:hep-th/9711067].
}

\lref\dkm{
A.~Dabholkar, R.~Kallosh and A.~Maloney,
{\it A Stringy Cloak for a Classical Singularity},
[arXiv:hep-th/0410076].
}

\lref\stvafa{
A.~Strominger and C.~Vafa,
{\it Microscopic Origin of the Bekenstein-Hawking Entropy},
Phys.\ Lett.\ B {\bf 379}, 99 (1996)
[arXiv:hep-th/9601029].
}

\lref\borde{
A.~Borde,
{\it Geodesic focusing, energy conditons, and singularities }
Class.\ Quantum.\ Grav.\ {\bf 4}, 343 (1987).}

\lref\horpol{
G.~T.~Horowitz and J.~Polchinski,
{\it A correspondence principle for black holes and strings},
Phys.\ Rev.\ D {\bf 55}, 6189 (1997)
[arXiv:hep-th/9612146].
}

\lref\wegotscooped{
A.~Sen, 
{\it How Does a Fundamental String Stretch its Horizon?},
[arXiv:hep-th/0411255].
}

\lref\SabraDH{
W.~A.~Sabra,
{\it Black holes in N = 2 supergravity theories and harmonic functions},
Nucl.\ Phys.\ B {\bf 510}, 247 (1998)
[arXiv:hep-th/9704147].
}

\lref\SabraKQ{
W.~A.~Sabra,
{\it General static N = 2 black holes},
Mod.\ Phys.\ Lett.\ A {\bf 12}, 2585 (1997)
[arXiv:hep-th/9703101].
}


%
\baselineskip 16pt \Title{\vbox{\baselineskip12pt
\line{\hfil UCB-PTH-04/34} 
\line{\hfil LBNL-56671}
\line{\hfil NSF-KITP-04-129}
\line{\hfil SLAC-PUB-10879} }}
{\vbox{
{\centerline{String-Corrected Black Holes} }}}
\centerline{\ticp Veronika E. Hubeny$^a$,
Alexander Maloney$^b$, 
and  Mukund Rangamani$^a$
\footnote{}{\ttsmall
 veronika@berkeley.edu, maloney@slac.stanford.edu,
mukund@socrates.berkeley.edu}}
\bigskip \centerline{\it $^a$
Department of Physics, University of California, Berkeley, CA
94720, USA} \centerline{\it  Theoretical Physics Group, LBNL,
Berkeley, CA 94720, USA}
\centerline{\it KITP, University of California, Santa Barbara, CA
93105, USA}
\bigskip \centerline{\it $^b$
SLAC and Department of Physics, Stanford University, Stanford, CA 94309}

\bigskip
\centerline{\bf Abstract}
\bigskip

We investigate the geometry of four dimensional black hole solutions in 
the presence of stringy higher curvature corrections to the low 
energy effective action.  For certain supersymmetric two charge black 
holes these corrections
drastically alter the causal structure of the solution, converting
seemingly pathological null singularities into
timelike singularities hidden behind a finite area horizon.
We establish, analytically and numerically, that the string-corrected
two-charge black hole
metric has the same Penrose diagram as the extremal four-charge
black hole. 
The higher derivative terms lead to another dramatic effect --
the gravitational force exerted by a black hole on
an inertial observer is no longer purely
attractive! The magnitude of this effect is related to the size of the
compactification manifold.

\smallskip 

\Date{November 2004}

\vfill \eject


\listtoc
\writetoc

\newsec{Introduction}

The physics of black holes has revealed many fascinating
insights into the workings of quantum gravity.
One of the most important observations, due to Bekenstein and 
Hawking, is that a black hole should be assigned an entropy 
equal to one quarter its horizon area,
\eqn\bhrel{
S=A/4.}
Strominger and Vafa showed that, for certain supersymmetric black 
holes, this entropy may derived microscopically in string theory
\stvafa.
They considered black holes with 
regular event horizons, whose areas are large in Planck units.
This insight followed the remarkable paper of Sen \sen,
who studied four dimensional black hole solutions of heterotic string theory 
that classically have singular event horizons with zero area.  The 
entropy of such black holes, which may be counted microscopically in
string theory, is nevertheless non-zero.
Sen conjectured that in this case string theoretic 
corrections lead to 
a finite area horizon, and used a brick wall model to reproduce 
the microscopic entropy up to a numerical coefficient.
This idea, that quantum corrections should render sensible 
an apparently singular solution of general relativity has
a rich history in string theory.
In this paper we will consider a class of quantum corrections,
which induce higher curvature terms quadratic in the 
Riemann tensor in the low energy effective action.  
For a wide class of black holes, 
these higher curvature corrections
are sufficient to convert previously singular solutions into regular 
black holes with a finite area horizon.

Many authors have considered black hole solutions in the presence of 
higher curvature terms.  In general one does not expect the 
relation between entropy and area to be of the Bekenstein-Hawking form 
\bhrel, this being specific to the Einstein-Hilbert form of the Lagrangian.
For generic higher derivative actions, Wald \wald\ has provided
a formula for the macroscopic black hole entropy,
generalizing the Bekenstein-Hawking relation \bhrel.
In a beautiful series of papers, the authors of \refs{\cdm, \cdkm}
discussed supersymmetric 
black hole solutions in the presence of higher curvature terms.
They showed that, for a class of black hole solutions including
those of \stvafa, this entropy formula correctly reproduces the subleading 
corrections to the microscopic entropy.\foot{
This result  is all the more remarkable given that Wald's 
original derivation applies only to non-supersymmetric black holes.  
Wald's prescription assumes the existence of a bifurcate Killing
horizon.  Moreover, it is based on the first law of thermodynamics, 
which assumes a finite temperature.  Neither of these
conditions holds for supersymmetric black holes, so in this case
Wald's formula must be regarded as a conjecture. However, given it's
 success in reproducing microscopic entropy formulae, it is an 
extremely well motivated one!
} 

Recently, Dabholkar \atish\ showed that these higher curvature
corrections have a profound effect on the physics
of the zero-horizon black holes considered by Sen.
Remarkably, in this case one can reproduce an infinite 
series of subleading corrections to the entropy.
Inspired by this, \dkm\ showed that, for broad class of
solutions that classically have zero horizon area, the 
area becomes non-zero once quantum corrections are taken into
account.  Thus quantum effects convert an 
apparently pathological solution of General 
Relativity into a regular black hole with an event horizon.
For these black holes, higher curvature corrections to
the Bekenstein-Hawking relation \bhrel\ are large -- 
in this case, $ S = A / 2$, where $A$ is the quantum corrected area.

All of the above derivations rely on the supersymmetric attractor mechanism
\fks,  which states
that the moduli of the compactification manifold approach fixed
values at the horizon.  This ensures that the
entropy of a black hole depends only on the
charges and not on the asymptotic values of the moduli. Moreover, 
it allows one to study near-horizon properties of a
black hole without ever solving for an explicit metric.

In this note we will study the full geometry of these solutions, 
not just the near horizon behavior, 
and find the black hole metric including stringy higher-curvature  
corrections. We will consider the four dimensional effective theory 
found by compactifying type IIA
on a Calabi-Yau three-fold. The resulting low energy action
is $d=4$, $\CN = 2$ supergravity coupled to vector and hyper multiplets -- 
the number of such multiplets and the form of the couplings depends on the
choice of Calabi-Yau.  
The four dimensional supersymmetric black holes of this theory correspond to
configurations of branes wrapping various cycles in the internal Calabi-Yau.
The metric can then be found by solving a BPS equation, whose form in
the presence of higher curvature corrections is quite complicated
\cdkm.

The prototypical example of this set-up is type IIA string
theory compactified on \kt. In this case one can construct
two-charge black holes by wrapping $Q_4$ D4-branes on ${
K3}$, along with $Q_0$ D0-branes. This geometry is U-dual to the
heterotic string  black hole considered by \sen, and also to the
standard D1-D5 system. The microstates of this system may be 
counted exactly, leading to
an entropy $S = 4 \, \pi \, \sqrt{Q_0 \, Q_4}$.

In the absence of higher curvature corrections, 
the four dimensional metric is
\eqn\dzdf{
ds^2 = -e^{2g} \, dt^2 + \, e^{-2g}\,  \(dr^2  + r^2 \, d\Omega_2^2\) \ ,
}
where
\eqn\asd{
e^{-2g} = \sqrt{\left(1 + {Q_0 \over r}\right)\left(1 + {Q_4 \over r}\right)}
\ . }
Near $r =0$, the metric becomes 
\eqn\nearhor{
ds^2 = - \rho^2 \, dt^2 + Q_0 \, Q_4 \, \(4 \, d\rho^2 + \rho^2 
\, d\Omega_2^2 \) \ ,}
where $r = \rho^2 \sqrt{Q_0 Q_4}$.  This solution is 
singular at $r= \rho = 0$, and has no event horizon.
This is to be contrasted with the usual extremal Reissner-Nordstrom 
solution, which is of the form \dzdf\ with $e^{-2g}=(1+{Q/r})^2$.  This
solution has a finite horizon, with area $4\pi\, Q^2$, and a near
horizon $AdS_2\times S^2$ geometry. Typically, 
black holes with four or more charges are of the Reissner-Nordstrom
type, and have a regular event horizon with finite area.

In string theory, both $\alpha'$ and $g_s$ effects will lead to
higher derivative terms in the low energy action.
For type IIA strings on a Calabi-Yau, this includes terms of the form
$F_g \, R^2 \, W^{2 g -2}$, 
where $W$ is the graviphoton field strength and the
$F_g$ depend on the Kahler moduli  of the Calabi-Yau. 
In string perturbation theory, such couplings arise at the
$g$-loop level.
In the language of $\CN=2$ supersymmetry, these couplings are 
F-terms, which can be efficiently computed in topological string theory --
in fact, $F_g$ is 
simply the $g$-loop topological string partition function.
This led the authors of \osv\ to conjecture a beautiful
relation between topological string and black hole partition functions
(see also \vafacy).

We will consider cases where the internal Calabi-Yau is 
large in string units, so higher order $\alpha'$ effects may be 
neglected.  In this case the tree and loop level terms involving
$F_0$ and $F_1$ are simple enough that the attractor equation can be
solved explicitly.
For \kt\ the higher loop contributions $F_{g>1}$ vanish, although 
in general such terms will be present.  In any case, 
once one has solved the attractor equations the metric is completely
determined by a Killing spinor equation.
As we will describe, this equation is a 
non-linear second order differential equation, which is generally difficult to
solve. Nevertheless, we can learn from this equation many
important features of the geometry. Moreover, it can be solved numerically.
This allows us to incorporate the one-loop $g_s$ corrections to the 
black hole solutions in a controlled fashion. 

We will show that once the higher curvature terms at one-loop are accounted for, 
the two-charge black hole is no longer described by the singular solution
\dzdf, \asd. Instead, it is given by a regular black hole solution,
with a timelike singularity hidden behind a finite-area event horizon.
In many cases, the curvature of this horizon is large, so further $g_s$ and
$\alpha'$ corrections may be neglected.
The causal structure of the solution is then 
identical to that of the four charge 
supersymmetric black hole spacetime.  The near-horizon geometry 
thus becomes $AdS_2 \times S^2$.
Apart from these results, which might have been expected from
the analysis of \dkm, we find some curious features.
The metric component $g_{tt}$ is no longer 
a monotonic function of radius $r$. Instead,
there are characteristic oscillations, whose 
period depends on the volume of some cycles in the internal
Calabi-Yau. This implies 
that the gravitational potential is no longer purely attractive!  
An inertial observer could stay perched at a local minimum
of the gravitational potential a finite, fixed 
distance from the black hole. This bizarre feature is due to the presence
of higher derivative terms in the action which violate the energy 
conditions; in particular, we will see that the null energy condition 
is violated. It must be emphasized that there is a version of the positive 
energy theorem (PET) for these spacetimes by virtue of supersymmetry. 
Thus the total energy is
positive definite, despite the 
presence of local pockets of negative energy density.

We begin in Section 2 with a brief summary of the generalized 
attractor mechanism in the presence of higher curvature 
interactions, as described by \refs{\cdm,\cdkm}. In Section 3 we 
analyze the Killing spinor equations for a class of Calabi-Yau 
compactifications, and infer some general properties of the resulting
black hole solutions. In Section 4 we study in detail the 
two charge supersymmetric black holes,  before concluding in 
Section 5.
\foot{
While this paper was in preparation, reference \wegotscooped\ appeared,
which discusses several similar issues for the $K3\times T^2$ black
holes.}

\newsec{Review of Supersymmetric Attractor Mechanism}
In this section we review the construction of supersymmetric black
hole solutions, including higher derivative $R^2$ corrections, 
given by \refs{\cdkm}. This framework is
appropriate  for studying black holes geometries in $ d =4 $, 
$\CN =2$ ungauged supergravity. As we describe in section 2.2, the Killing
spinor  equation for the metric in this case a second order
differential equation.  In section 2.3 we study this equation for
Calabi-Yau compactifications of type IIA string theory, including tree
level and one-loop contributions to the effective action.  We discuss
the special case of \kt\ in 2.4.

\subsec{Basic Setup}
Before delving into details, we will first review the construction
of $\CN=2$ supergravity actions with higher derivative 
terms.  We will discuss here only a few salient details -- a more
complete review may be found in \mohaupt, and references therein.

We start with a gauged $\CN=2$ superconformal theory
in four dimensions. The field content of the theory, known as the Weyl
multiplet, contains one gauge field for each generator of the
$SO(4,2)$ superconformal group.  This includes a veirbein and spin
connection, which gauge translations and Lorentz transformations,
as well as other fields that gauge dilatations, special conformal
transformations and the various fermionic symmetries. 
This superconformal theory has the advantage that higher
curvature terms are relatively easy to construct. 
To relate this to standard $\CN =2$ supergravity, it is necessary to gauge fix
the superconformal symmetry down to $\CN=2$ Poincar\'e
supersymmetry. This is achieved by the introduction of 
an extra multiplet, known as the
conformal compensator, whose details will not be important for the
current discussion. We will not review the details of the gauged 
superconformal theory, but instead simply quote the results as needed.

The low energy theory also includes $n_v+1$ vector multiplets, 
labeled by an index $I=0, \cdots ,n_v$. One linear combination of these 
vector fields is the graviphoton, and the others are associated with
deformations of the internal Calabi-Yau. 
The Weyl multiplet does not include the graviphoton, which
gauges the central $U(1)$ of the theory, so we have included 
it explicitly as a vector multiplet. 
The lowest components of the vector multiplets are $n_v+1$ complex scalars
$X^I$.
In type IIA string theory, the $X^I$ are projective
coordinates parameterizing the Kahler moduli space of an internal Calabi-Yau. 
The theory also includes hypermultiplets, which are related to 
complex structure deformations of the Calabi-Yau, but these will not 
be important for our discussion.

The couplings between the vector and Weyl multiplets 
are summarized by a prepotential $F(X^I,W^2)$, where
$W^2$ is the square of the graviphoton field strength.  
The function $F$ is holomorphic in the
$X^I$ and obeys the homogeneity condition 
\eqn\asd{
2\,  X^I\,  F_I + W^2 \, F_{W^2} = F \ .}
Here $F_I$ and $F_{W^2}$ denote partial derivatives of
$F$ with respect to $X^I$ and $W^2$, respectively.  

When $F$ is
independent of $W^2$, this is the usual holomorphic
prepotential. When  $F$ contains a  term linear in $W^2$, 
the action includes curvature squared interactions.
In general, we may expand the prepotential as
\eqn\prepot{
F(X^I, W^2) = \sum_{g=0}^{\infty} \, F_g(X^I) \, W^{2g} .}
In IIA string perturbation theory, the $g^{th}$ term in this expansion 
appears as the $g$-loop level.  To see this, recall that the 
dilaton, which determines the string coupling, lives in a hypermultiplet.
The coupling between vector and hyper multiplets is highly 
constrained in $\CN =2 $ theories, completely fixing the dependence
of the terms in \prepot\ on $g_s$.  
The coefficients $F_g(X^I)$ are most readily computed in terms of 
a topological string partition function. 
We will write down a few explicit prepotentials in sections 2.3 and 2.4.

\subsec{Black Hole Solutions}

We 
will follow the notation used in \cdkm, who discussed attractor
and Killing spinor equations for
stationary multi-center solutions. We will consider here only
static, spherically symmetric black hole solutions of the form
\eqn\statmet{
ds^2 = -e^{2 \, g} \, dt^2 + e^{-2 \,g} \, \(dr^2 + r^2 \, d\Omega_2^2 \)
.}
The electric and magnetic charges of the solution are denoted
$p^I$ and $q_I$, where $I=0,..., n_v$.
The solution depends on these charges via the harmonic functions
\eqn\harmonic{
H^I(r) = h^I + {p^I \over r} \ , \qquad H_I(r) = h_I + {q_I \over r} \ ,}
where $h^I$ and $h_J$ are constants related to the asymptotic values of the 
moduli $X^I$.

The moduli $X^I$ will in general depend on $r$, and are
determined in terms of the harmonic functions \harmonic\ by 
a set of stabilization equations \refs{\SabraDH,\SabraKQ}.  
In terms of the rescaled variables
\eqn\asd{\eqalign{
Y^I = e^{K/2} {\bar Z} X^I,&~~~~~ \Up= e^K {\bar Z}^2 W^2 \cr
e^{-K} = i({\bar X^I}F_I - {\bar F}_I X^I),&~~~~~
Z = e^{K/2} (p^I F_I - q_I X^I ))
}}
the stabilization equations are
\eqn\genstab{\eqalign{
Y^I - &\bY^I = i \, H^I  \cr
F_I(Y,\Up)  - &\bF_I(\bY,\bUp) =  i\,  H_I
.}}
In addition, the graviphoton field strength is fixed to be
\eqn\gravph{
\Up = -64 \, \(\nabla_p \, g\)^2.
}

Having fixed the moduli, the metric is determined
by a BPS equation, which takes the form \cdkm\
\foot{The authors of \cdkm\ included a constant $\chi$ in their 
discussion of the BPS equations, which 
sets the normalization of the Einstein-Hilbert 
term in the Lagrangian.  We have set here $\chi = -2$.} 
\eqn\metdet{
i\, \[\bY^I \, F_I(Y,\Up) - \bF_I(\bY,\bUp) \, Y^I\] -
 \, e^{-2 \,g } = 128 \, i \, e^g \, \nabla^p \[ \, \(\nabla_p
e^{-g} \) \, \(F_\Up - \bF_\Up \) \, \] \ . }
The indices $p,q$ label the three flat spatial directions.

\subsec{Calabi-Yau Black Holes}

For type IIA string theory on a generic Calabi-Yau, 
the number of vector multiplets $n_v$ is fixed by the 
topology of the Calabi-Yau -- roughly, it counts the number of
four-cycles in the Calabi-Yau. The moduli $Y^I$ then measure 
the sizes of these cycles.  When the Calabi-Yau is
large, so that $\alpha'$ corrections may be neglected, the first 
two terms in the pre-potential are
\eqn\prepotcy{
F(Y^I, \Up) = D_{ABC}\, {Y^A \, Y^B\, Y^C \over Y^0} +
{ d_A \, Y^A \over Y^0} \, \Up\ .
}
Here $D_{ABC}$ is related to the intersection numbers
and $d_A$ is given by the second Chern class of the
Calabi-Yau dotted into the various four-cycles.
The constraint that $\alpha'$ corrections are small requires that 
certain ratios of the charges be taken large.  

For this prepotential
the attractor equations \genstab\ can be solved exactly if we take 
$H^0=0$. To this end, we 
define $D_{AB} = D_{ABC} \, H^C$ and its inverse 
$D_{AB} \, D^{BC} = \delta_A^C$. Then \foot{We are using the 
fact that \gravph\ implies that $\Up$ is real. For stationary 
solutions this is no longer guaranteed.}
\eqn\gensol{\eqalign{
Y^A &= {i\over 2} \, H^A + {1\over 6} \, Y^0 \, D^{AB} \, H_B \ , \cr
Y^0 &= \sqrt{{ {1\over 4} \, D  -  d_A \, H^A \Up \over H_0 + {1\over 12}
\, D^{AB}\, H_A \, H_B }} \ 
}} 
where 
\eqn\asd{
D = D_{ABC} \, H^A\, H^B \, H^C \ .}
When $D_{AB}$ is not invertible, the matrix $D^{AB}$ in these 
expressions is replaced by the pseudo-inverse of $D_{AB}$.

From \gensol, it follows that
\eqn\varquan{\eqalign{
i \, \[ \bY^I \, F_I - Y^I \, \bF_I \] &= -{
D \over Y^0 } + 2 \, \Up \, {d_A \, H^A \over Y^0}
\cr
F_\Up -\bF_\Up & = i \, {d_A \, H^A \over Y^0}
.}}

If we define
\eqn\notation{\eqalign{
a(r) & = H_0 + {1\over 12} \,H_A \, D^{AB} \, H_B \ , \cr
b(r) & = 64 \, d_A \, H^A \ , \cr
d(r) &= D =  D_{ABC} \,H^A \, H^B \, H^C \ , 
}}
then, using \gravph, and \gensol, we find
\eqn\yzexp{
Y^0 = {\sqrt{ d(r)  + 4 \,g'(r)^2 \, b(r) \over 4 \, a(r)}}\ .} 
In addition, introducing 
\eqn\xidef{
\xi(r) = {b(r) \over  Y^0} \ , }
we can write \metdet\ as 
\eqn\kill{
e^{-2 \, g(r)} = {d(r) \over Y^0} + 
2 \, \[ \xi(r) \, \(g''(r) + {2 \over r} \, g'(r)  \) +
\xi'(r) \, g'(r) \] \ .}
While we have reduced the problem of finding the metric to solving
a differential equation, this is still a fairly complicated non-linear problem.

\subsec{The \kt\ Example}

Let us now consider a special example, where the Calabi-Yau is 
\kt.  In this case the resulting low energy theory actually has
$\CN=4$ supersymmetry, but we will continue to use the 
use the $\CN =2$ language. There are now 24 vector multiplets, 
corresponding to the two cycle of $T^2$, the 22
two-cycles of {K3}, and the graviphoton.
The prepotential takes the form
\eqn\lvprepot{
F(Y^I,\Up) = -\, \( { Y^1 \over Y^0} \) \, 
\(  {1\over 2} \, C_{\a \b} \, Y^\a \, Y^\b \, +{ 1\over 64} \, \Up\) \ 
} 
where $\a,\b = 2, \cdots 23$ label the two-cycles of {K3}
and $C_{\a \b}$ is the intersection matrix of {K3}.
For a general Calabi-Yau there will be loop corrections
of order $\Up^2$ and higher to the prepotential.  
However, for \kt\ these additional
contributions vanish and the prepotential has only tree-level
and one loop contributions.
This full, quantum corrected prepotential can be found exactly
(see {\it e.g.} Eq. (19) of \atish), although we will focus only on terms 
that are tree level in $\alpha'$, which are given by \lvprepot.

We are interested in black hole solutions 
corresponding to the D0-D4 systems discussed in the
introduction.  For these black holes, it is sufficient to take
\eqn\hdefs{
H^1 = 1 + {Q_4 \over r} \ , \qquad H_0 = 1 + {Q_0 \over r} \ , \qquad
H^a = h^a \ , \qquad H_a = H^0 =  0 \ . 
}
If we define
\eqn\zetadef{\eqalign{
Y^0 &= \sqrt{{\b \over 4} + g'^2} \, \sqrt{{H^1 \over H_0}} \ , \cr
\b &= {1 \over 2} \, C_{a b} \, h^a \, h^b \ , \cr
\zeta(r) &= {H^1 \over Y^0} = \sqrt{H^1 \, H_0}\, {1 \over \sqrt{{\b \over 4}
+ g'^2}}  \ ,
}}
then the equation of motion for the metric is
\eqn\final{
 e^{- 2 \, g} =  \beta \, \zeta + 2 \, \[ \zeta \(g''+ 2 \,
{g' \over r} \,\) + \zeta' \, g' \] \ . }
Solutions to \final\ then describe $g_s$-corrected black hole 
geometries.

\newsec{Black Hole Solutions: General Structure}

In this section we study the solutions to the Killing spinor equation
\kill.  The solutions are difficult to find analytically,
but we will see that nevertheless one can extract the causal
structure.  We will consider a broad class of Calabi-Yau black holes,
which classically (\ie, without $R^2$ corrections) have null singularities
and zero horizon area.  We show that once higher curvature corrections
are added, the horizon becomes non-singular and the causal structure is 
identical to that of the familiar extremal Reissner-Nordstrom solution.

\subsec{Leading Order Solutions}

First, we review the solutions in the absence of $R^2$ corrections.  
The $g''(r)$ and $g'(r)$ terms drop out of the Killing spinor equation 
\kill, and the warp factor $e^{-2 \,g}$ in the metric is
determined algebraically:
\eqn\leading{
e^{-2\, g} = 2 \, \sqrt{ d(r) \, a(r) } \ . }
At infinity the solution is asymptotically flat, with
\eqn\leadasym{
e^{2\, g} \to \; 1 \ , \qquad r \to \; \infty \ , }
provided we adjust the asymptotic values of the moduli $(h^A,h_A)$ 
appropriately.  When a sufficient number of magnetic
charges $p^A$ are non zero, so that 
$D_{ABC}\, p^A\, p^B\, p^C \ne 0 $, the solution has the form
\eqn\leadzero{
e^{2\, g} \to \; \( {r\over R_0} \)^2 \ , \qquad r\to \;0 \
}
near the origin.  Here
\eqn\ris{
R_0 = \sqrt{D_{ABC}\, p^A\, p^B \, p^C \, {\hat q_0}} }
where ${\hat q}_0 = q_0 + {1\over 12 } \, {\tilde D}^{AB} \, q_A \, q_B$,
with ${\tilde D}^{AB}$ the (pseudo) inverse of $D_{ABC} \, p^C$.
When $R_0\ne0$ the geometry has a non-singular
event horizon  of area $4\, \pi \, R_0^2$, and the near horizon
geometry is $AdS_2\times S^2$.

\ifig\penrdnull{Penrose diagram for the classical black hole with vanishing 
horizon area.}
{\epsfxsize=5cm \epsfysize=5cm \epsfbox{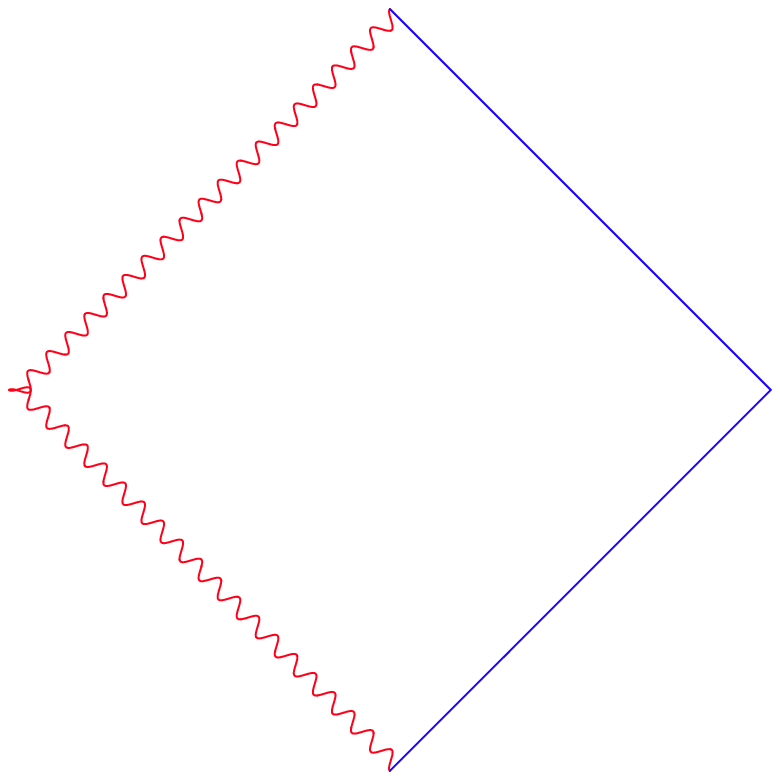}}

For the rest of this paper, we will consider solutions with 
$D_{ABC}\, p^A \, p^B \, p^C=0$. In this case the radius
\ris\ vanishes and the function $e^{2g}$ vanishes more slowly as
$r\to0$.  In particular, 
\eqn\leadzeroa{
e^{2\, g} \sim r^\alpha \ ,\qquad  r \to \; 0 \ ,
}
where $\alpha$ is ${1\over 2} $, $1$, or ${3 \over 2}$ depending on
the number of
non-vanishing magnetic charges.  For these solutions the
curvature diverges as $r\to 0$. 
For the D0-D4 system on \kt, $\alpha=1$, as can be seen from \dzdf. 
In any case, the curvature singularity is null, and coincides with 
the horizon at $r=0$.
The Penrose diagram for these black holes is shown in \penrdnull.
We will now move on to the $R^2$ corrected metric.

\subsec{Asymptotic Behavior}

Before delving into details, we will first consider the asymptotic 
behavior of the solutions to the Killing spinor equation \kill.

At $r\to\infty$, the differential equation \kill\ simplifies considerably.
We have
\eqn\asyma{
e^{-2\, g} \to 2 \, \sqrt{ a(r) \, d(r)} \ , }
plus corrections at $\CO\(r^{-3}\)$. This goes to a constant
out at infinity, indicating that the space is asymptotically flat. 
Moreover, by an appropriate choice of  the asymptotic values of the 
moduli $(h_A,h^A)$ we can ensure that $e^{2 \, g } \to 1$. 
Of course, this is provided that $D_{ABC}\, h^A \, h^B \, h^C$ and
$h_0 + {1\over 12} D^{AB} \, h_A \, h_B$ are non-zero. As is clear 
from \asyma, we will find a non-asymptotically flat geometry when either of
these quantities vanish.
For instance, consider the simple case
\eqn\degenex{
H^A \equiv 0 \ , \forall \, a \neq 1 \ , \qquad {\rm and} 
\qquad H^1 \neq 0 \ .}
Then from \notation\ and \yzexp\ we find that 
\eqn\specyz{
d(r) = 0 \ , \qquad {\rm and} \qquad Y^0 = g'(r) \sqrt{{b(r) \over a(r)}}.}
It is easy to show that \kill\ reduces for this special case to 
\eqn\algkill{
e^{-2\, g} = 2 \, \( {d \over dr} \, \sqrt{a(r) \, b(r) } 
+ {2 \over r} \sqrt{ a(r) \, b(r)} \) \ . }
This implies that $e^{-2 \, g} \to c/r$, with 
$c = 4 \sqrt{a_\infty \, b_\infty}$, for large $r$. The metric
is not asymptotically flat. 
\foot{ 
Strangely enough, the asymptotic metric in this case looks like
the singular near-horizon metric \nearhor. 
We should remark here that there is no singularity in this case. The 
geometry \nearhor\ is singular only as $ r \to 0$; for large $r$ it is 
a geodesically complete spacetime.}
To understand this, note that \degenex\ implies that at large $r$
we are driven towards a singular point in the 
Calabi-Yau moduli space.
In particular, as $r\to \infty$ some of the ratios
$Y^A/Y^0$, which measure sizes of 
cycles in the internal Calabi-Yau, become small.  Thus $\a'$ effects
can no longer be neglected. This effect 
is seen from the four dimensional perspective as a non-asymptotically 
flat geometry. For the rest of the paper we will ensure that 
this does not happen by choosing constants $(h^A,h_A)$ 
such that $a(r)$ and $d(r)$ are non-zero at large $r$.

Let us now consider the metric at $r\to 0$.
Once higher curvature corrections are taken into account,
\eqn\zeroa{
e^{2g} \to \; \left({r\over R_0}\right)^2 \ ,  \qquad r\to \; 0 \ ,
}
so the near horizon geometry is $AdS_2\times S^2$,  with finite radius $R_0$.
To see this, note that with the ansatz \zeroa\  the Killing spinor 
equation \kill\ becomes
\eqn\zerob{
e^{-2g} = {2 \over r^2} \, \sqrt{d_A p^A  {\hat q}_0}
}
plus sub-leading corrections of order $\CO(r)$.
\foot{We have used the fact that,
since we are considering solutions with $D_{ABC}\, p^A \, p^B \, p^C=0$, 
the function $D/Y^0$ may be neglected as $r \to 0$. 
}  
This implies that the solution takes the form \zeroa, provided
\eqn\adsrad{
R_0 = \sqrt{2} \(d_A \,p^A \,{\hat q}_0 \)^{1/4}
}
is non-zero.  This corresponds to a horizon area 
\eqn\area{
A = 4 \, \pi\, e^{-2 \,g(r)} \, r^2 |_{r=0}= 
8 \, \pi \, \sqrt{ d_A \,p^A \,{\hat q}_0}   \ , }
which agrees with the result of \dkm. 
Note that we could have concluded this earlier, from \algkill. 
Even though the choice of moduli \degenex\ leads to problems 
in the asymptotic form of the metric, the supersymmetric
attractor mechanism guarantees that the near horizon physics
is independent of this asymptotic behavior.
Typically, by taking $q_0, p^A \gg 1$ this
area can be made macroscopic, \ie much larger than the
string or Planck scale.
\foot{
We should emphasize that this holds only for the generic case -- 
for $K3\times T^2$, the string coupling is determined by the
charges (due to the enhanced $\CN=4$ supersymmetry).  In this case, 
the resulting horizon is string scale.} 

\subsec{Horizons and Singularities}

Having understood the asymptotic behavior of the solution, we can now
ask what happens in the intermediate region where $r$ is finite and
non-zero.  The Killing spinor equation is quite difficult to solve
analytically, but we may nevertheless establish several results concerning
the causal structure of the solution.

First, we would like to ask whether the $R^2$ corrections lead to
additional singularities or horizons at finite $r$.  In the parameterization
\statmet, these correspond to points where $e^{2g}$ diverges or vanishes
at finite values of $r\ne 0$.  We will now demonstrate that this is
not the case, and that the metric is smooth and non-degenerate for
all positive values of $r$.

Let us assume was have such a point, where $|g|\to \infty$ at a finite,
positive value of $r=r_o$. At this point, both $|g'(r)|$
and $Y_0 \sim |g'(r)|$ diverge.  The Killing equation \kill\
then implies that
\eqn\asd{
e^{-2g} = {1 \over 8} \,  \( {2\over r} + \p_r \) \, 
\sqrt{ b(r) \, a(r) } \ .}
It is straightforward to see that
the right hand side is finite and non-zero for any
finite $r_o$.  This contradicts our assumption that $|g|\to\infty$,
so we conclude that the solution is non-singular and non-degenerate.

\newsec{Black Holes Solutions: Detailed Analysis}

We now turn to a detailed analysis of \kill, which as we have seen is 
difficult to treat analytically. \foot{One could consider series solutions 
to \kill; however, expanding $e^{-2 \, g}$ in a power series 
around the origin or around $ r = \infty$, we find that successive series terms tend to 
lower the radius of convergence. One possibility is to 
resum the resulting series and analytically continue beyond the
domain of convergence, obtaining a closed form 
solution. Unfortunately, this method proves too complicated, 
so  we turn to numerical solutions.}
We will see that the numerical solutions provide a good 
deal of information about the spacetime and are quite robust.

\ifig\fqzqf{Plot of the function $e^{2 \,g}$ for $Q_0 = 500$ and $Q_4 = 60$.
The first plot shows the behavior of the function for small $r$, and 
the second plot the behavior at large $r$.}
{\epsfxsize=15cm \epsfysize=4.5cm \epsfbox{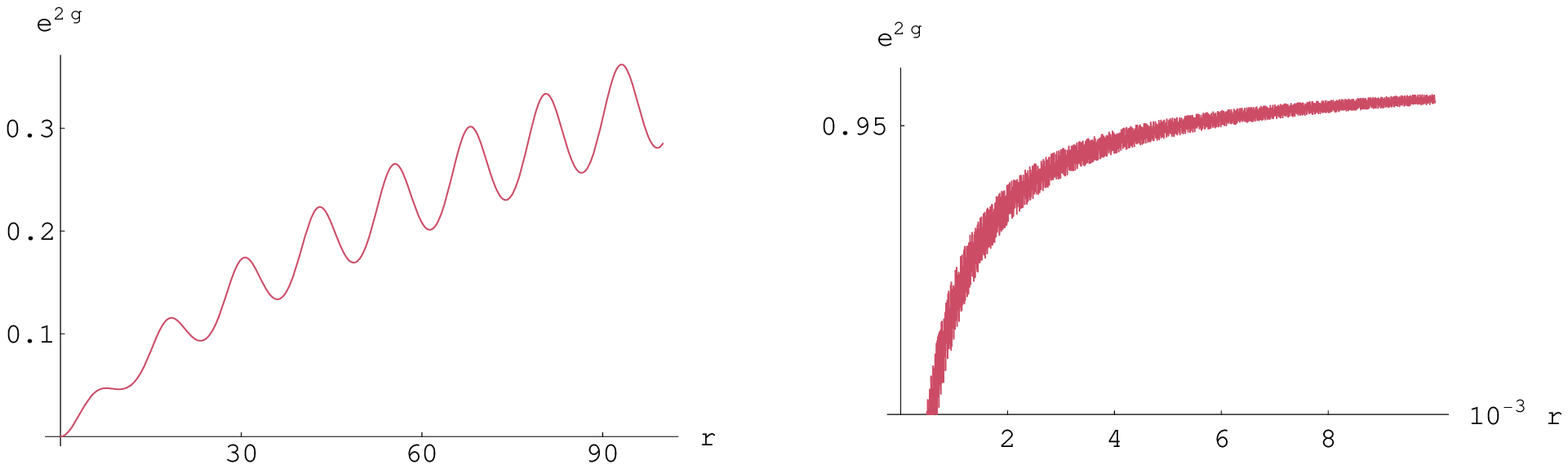}}

\subsec{Numerical Solutions}

In this section we describe numerical solutions to the Killing spinor
equation \kill, which verify the analytic arguments of the previous section.
For simplicity, we will consider the case where all but one of the
magnetic charges vanishes.  This includes the two-charge black hole 
configurations on \kt, mentioned earlier and the subject of \atish.

In this case, one can find a constant $\gamma$ such that
\eqn\dsimple{
d(r) = \gamma^2 \, d_A \, H^A \;\;\; \Rightarrow \;\;\;\; Y^0 = 
\sqrt{{\gamma^2 \over 256} + g'^2} \, \sqrt{b(r) \over a(r)} \ , }
and the equations of motion simplify considerably. In fact, it is 
easy to see that \kill\ can be written in the form \final, 
with $\gamma^2 = 64 \, \beta$. We will henceforth 
concentrate on equation \final, with the various parameters as 
defined in \hdefs, \zetadef. This allows us to make contact 
with \dzdf\ at various stages of the analysis.

\bigskip
\ifig\fnfull{Plot of the function $e^{2 \,g}$ for $Q_0 = 500$ and $Q_4 = 60$.
The divergence of the function as $r \to -Q_4$ is indicative 
of a singularity.}
{\epsfxsize=9cm \epsfysize=5cm \epsfbox{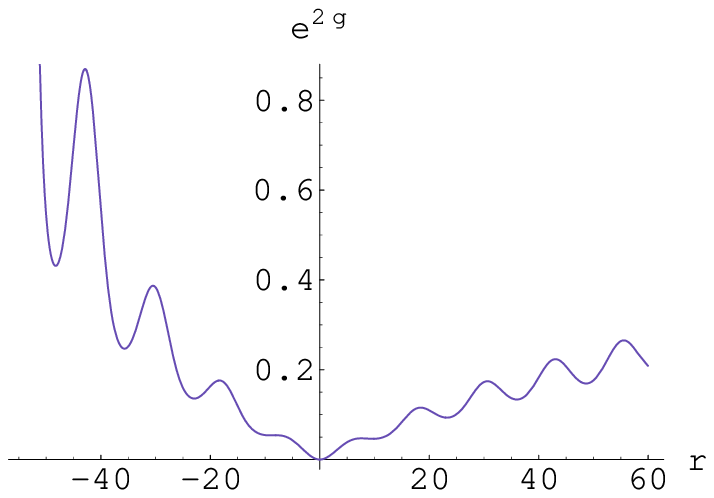}}

First, note that for an asymptotically flat solution, we need 
to choose $\beta = 1/4$. This ensures that the metric is correctly 
normalized at infinity, \ie, $e^{2 \, g} \to 1$. 
In \fqzqf, we show the numerical solution for 
the case $Q_0 = 500$ and $Q_4 = 60$. 
In \fnfull, we extend the same solution to negative values of $r$ -- this
is the region behind the horizon.  
One can find such solutions for 
any choice of $(Q_0, Q_4, \beta)$. In determining the boundary conditions 
for the numerical evolution we make use of the series solution around 
$r=0$ to set the value of the $g(r)$ and $g'(r)$. As can be seen from the 
plots \fqzqf, this solution satisfies the required near-horizon and asymptotic 
behavior, {\it viz.},  AdS$_2 \times S^2$ and asymptotically flat
(since the oscillations die out at infinity).

\subsec{Non-Monotonic Metric Functions}

The most striking feature of the solutions \fqzqf\ and \fnfull\
is that the metric function $e^{2 \, g}$ is not monotonic. \foot{Curiously, 
integrating in from $ r = \infty$ with appropriate boundary conditions, 
$e^{2 \, g}$ is monotonic but encounters a singularity at some $r >0$. 
Since we are interested in solutions with a regular horizon we discard 
this solution as unphysical.} 
This is in marked contrast with the leading order solution \dzdf.
One can check numerically that the period of the oscillations 
is independent of the charges.  In fact, the oscillation 
frequency $\omega$ is given by
\foot{
To understand this feature 
analytically, note that when $g'$ is small, the equation of motion
\final\ becomes
\eqn\asd{
\left(\p_r^2 + \beta\right)e^{2g} = \zeta^{-1}
.}
Thus in this regime the $e^{2g}$ takes the form of a 
homogeneous solution, which is periodic with frequency $\sqrt{\beta}$, plus
a non-periodic particular solution. 
}
\eqn\period{
\omega =  \sqrt{\beta}  \ , }
so that the separation between local maxima is $2 \, \pi/\sqrt{\beta}$.
Note that the characteristic oscillation period is set
by the volume of some cycle of the Calabi-Yau at large $r$. 
For example, it is easy to check that $Y^1/Y^0 \to 2/\sqrt{\beta}$ 
as $r \to \infty$, for this class of examples.  

The main effect of the non-monotone metric function is that there are 
inertial observers, who can sit at a fixed constant distance away 
from the black hole; the gravitational force felt by an inertial
observer will not be universally attractive towards the origin.
Instead, there will be small regions where the force is {\it repulsive}.
To see this, note that the function $e^{2g}$ is the gravitational
potential felt by a timelike geodesic.  In particular,
timelike radial geodesics, parameterized
by an affine parameter $s$, are given by
\eqn\timelike{
{\dot r}^2  = E^2 - e^{2 \, g} \ , \qquad t(s) = E \int^s e^{-2g(r(s'))} ds' ,
}
where $E$ is a constant of motion.  The dynamics are simply that
of a particle of energy $E^2$ moving in a potential $e^{2\, g}$.
An inertial observer may then sit at a finite value of $r$
where $e^{2\, g}$ is at a local minimum, where the attractive and repulsive
gravitational forces are balanced. 

\ifig\engcond{Plot of $R_{\mu \nu} \, \xi^\mu \, \xi^\nu$ to demonstrate 
failure of the null energy condition. One again we choose $Q_0 = 500$ and 
$Q_4 = 60$. $\xi^\mu$ is  chosen  to be the null vector tangent to 
radial null geodesics.}
{\epsfxsize=15cm \epsfysize=5cm \epsfbox{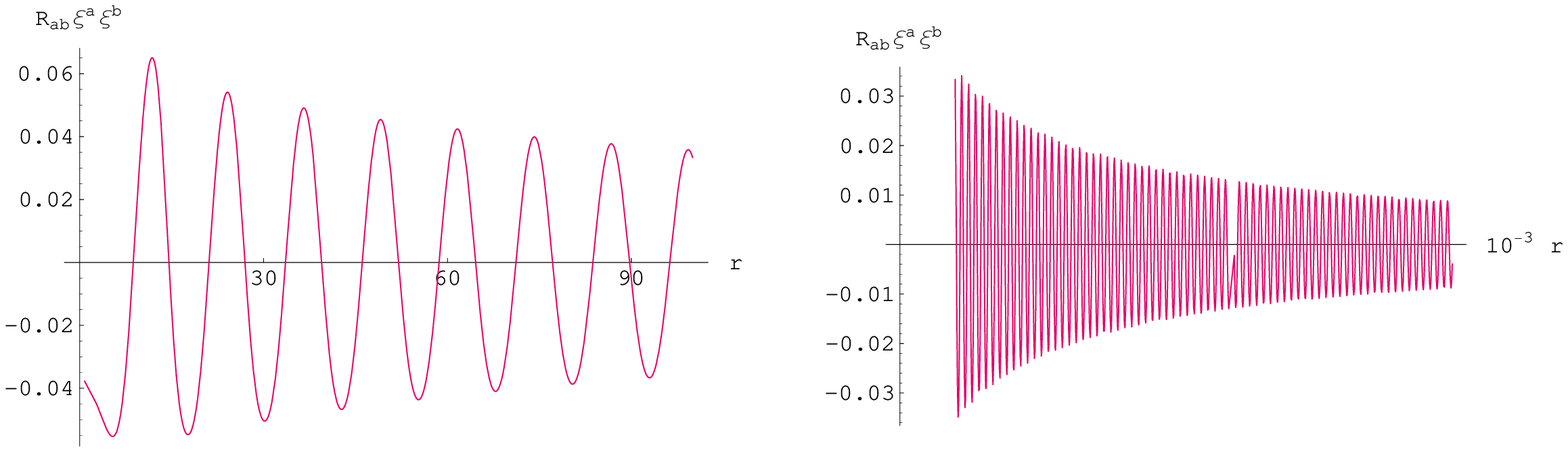}}

The reason for the bizarre behavior can be traced to the fact that
the $R^2$  terms violate energy conditions. In this event, it is
possible to have local  pockets of negative energy density, which will
manifest themselves in the form of repulsive gravitational
force. Given the numerical solution for the metric,  it is easy to
check this violation of energy conditions. Of course, since we have  a
higher derivative action, it is not clear how we should define the
stress tensor.  One option is to incorporate the higher derivative
terms as explicit contributions to the  stress tensor, writing the equation of
motion schematically as $G_{\mu \nu} = 8 \, \pi \, T^c_{\mu \nu}$,
where $T^c_{\mu \nu}$ includes  both the matter and the higher
derivative gravitational contributions.  With this definition
it is easy to define the energy conditions  in the
conventional fashion.  For example, the null energy condition will
require that $R_{\mu \nu } \, \xi^\mu \, \xi^\nu \ge 0$ for any null
vector $\xi^\mu$.  We show that this is not satisfied in the present
example, by exhibiting a null  vector for which
$R_{\mu \nu} \xi^\mu \, \xi^\nu$ is not positive definite in 
\engcond. It suffices to take $\xi^\mu$ to be the null vector tangent to radial 
null geodesics in the $(r,t)$ plane, \ie, $\xi = \p_t + f(r) \, \p_r$.
At the same time the supersymmetric nature of the solution guarantees
us that these backgrounds respect the positive energy theorem (PET).
This however only implies that the total energy of any excitation in
the system  is bounded from below by zero, with the supersymmetric
black hole  being the ground state. Since this is only a statement
about the total energy as measured by an asymptotic observer, there is
no contradiction with  the presence of local pockets of negative
energy.

We note in passing that, due to the  violation of the energy
conditions,  the null convergence condition does not obviously 
follow from the Raychaudhuri equation.  However, our spacetime 
satisfies the Averaged Null Energy Condition (ANEC), which requires 
that along every affinely parameterized inextendible null geodesic in 
the spacetime with tangent $\xi^a$ and affine parameter $\lambda$,
\eqn\anec{
\int \, d \lambda \, R_{\mu \nu} \, \xi^{\mu} \, \xi^{\nu} \ge 0 \ . 
}
It can be shown that the focusing theorems are satisfied 
with the weakened ANEC \borde. This, in particular, should 
ensure that the area theorem holds for our spacetime. 
In fact,  since the relation between entropy  and area is $ S =
A/2$ \dkm,  in order for the second law of thermodynamics to be
satisfied it is clear that the area theorem must be true.
  
\subsec{Causal Structure}

We will now investigate the causal structure of the higher-curvature
corrected solution. The causal structure is fixed completely by the
radial null geodesics, which follow trajectories $t(r)$ given by
\eqn\tint{
t^\pm(r) = \pm \int^r e^{-2g(r')} dr'
.}
Here $t^+$ and $t^-$ denote ingoing and outgoing geodesics.
Since the integrand is strictly positive and finite,
$t^\pm(r)$ are strictly monotone and continuous functions of $r$.
Moreover, we established that near $r=0$ and $r=\infty$
the form of $e^{-2g}$ is, up to overall constants, precisely that of
the four charge black hole.  These two results are sufficient to establish
that these two geometries have precisely the same causal structure
outside the horizon.

From the full numerical solution we can verify concretely that the 
causal structure of the string-corrected two charge black holes 
is identical to that of the four charge supersymmetric black hole. 
First of all, note that the metric function $e^{2 \, g}$  is positive 
definite and finite for $r > 0$. So the $(r,t)$ coordinates in 
\statmet\ are a good coordinate chart for $r >0$, with 
${\p \over \p t}$ being a timelike Killing vector.

\ifig\ricci{Plot of the Ricci scalar for $Q_0 = 500$ and $Q_4 = 60$.
The divergence as $r \to -Q_4$ indicates the location of the 
timelike singularity.}
{\epsfxsize=9cm \epsfysize=6cm \epsfbox{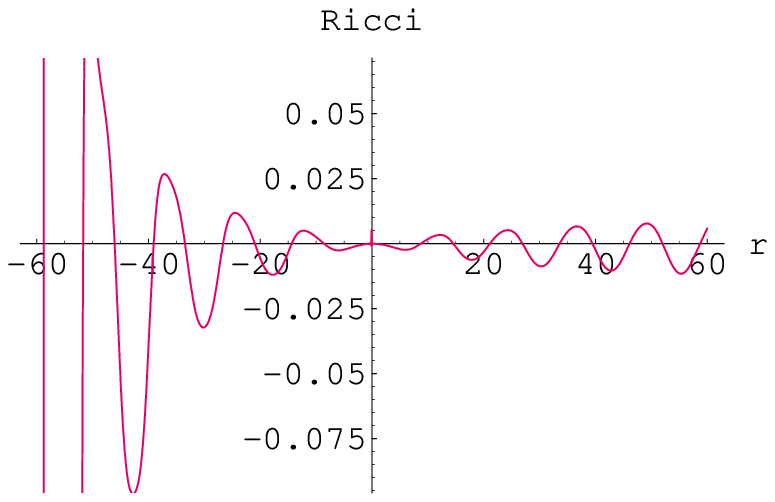}}

The coordinate chart breaks down at $r = 0$, where the Killing vector 
${\p \over \p t}$ becomes null. A non-degenerate 
coordinate chart can be introduced following the standard route for 
constructing ingoing/outgoing Eddington coordinates: 
$u,v = t \mp \int^r \, \; e^{-2 \, g(r')} \, dr' $. All we learn 
from the exercise is that the surface $ r = 0$ is a null surface, which 
can be thought of as the Poincare horizon of the Bertotti-Robinson 
spacetime. Since this is just a coordinate singularity, we can 
continue to the region $r < 0$. Once again we find that the 
function $e^{2 \, g}$ is positive definite, and hence ${\p \over \p t}$
is a timelike Killing vector. However, we find that (we are assuming that 
$Q_0 >>  Q_4$, as is required to suppress $\alpha'$ corrections 
to the pre-potential)
\eqn\singular{
e^{2 \, g} \to \infty \ , \qquad    r \to -Q_4 \ ,
}
implying that there is a singularity. One can indeed check this explicitly by 
considering the curvature invariants; for example, the Ricci scalar 
blows up as we approach $r = - Q_4$ as can be seen in \ricci.
\ifig\penrdcorr{Penrose diagram for the quantum corrected black hole.}
{\epsfxsize=3cm \epsfysize=6cm \epsfbox{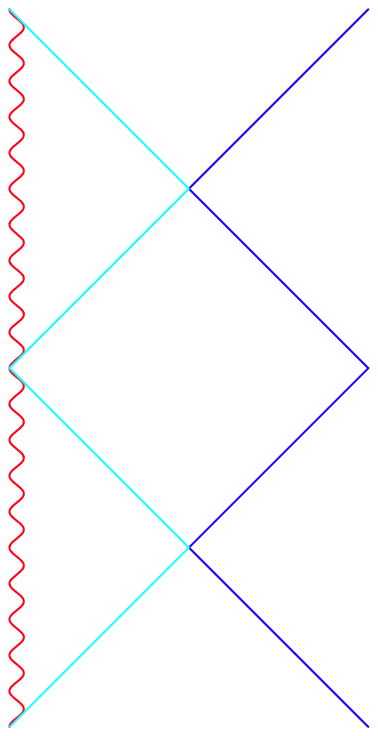}}
We conclude 
that the class of quantum corrections considered in this paper convert 
null singular spacetimes to regular black hole geometries.  
The Penrose diagram of this spacetime 
is therefore as shown in \penrdcorr.

\newsec{Discussion}

We have considered supersymmetric black holes in four dimensions
that arise in string theory compactified on a Calabi-Yau.  Our focus 
has been on solutions which seem to have 
singular event horizons, when viewed as solutions to the low 
energy effective $\CN =2 $ supergravity theory. 
Quantum effects, in the form of higher derivative 
corrections to the Einstein-Hilbert 
action, serve to smooth out the horizon to a regular 
null hypersurface. 
This result is valid for a large class of black holes 
which arise in Calabi-Yau compactifications. Our main result 
is that the causal structure of the quantum corrected solution is identical 
to that of the four charge supersymmetric black hole in four dimensions.

The quantum corrected metric exhibits a very unusual feature: the metric 
function is non-monotonic outside the horizon.
To the best of our knowledge, this is the 
first example of  a black hole with this feature. 
As a result, these black hole 
solutions admit inertial observers who can passively view the black hole 
from a fixed finite distance.  This is possible because local pockets of 
negative energy are allowed by virtue of higher derivative terms which
violate an energy condition. Furthermore, we have demonstrated 
that the spacetime satisfies the Averaged Null Energy Condition, which 
is sufficient to prove the area theorem in the present context.

At this point we should ask to what extent our approximations are
justified. We have 
incorporated the one-loop correction to the pre-potential, but 
have otherwise been working at tree level in $\a'$. 
This is justified when we can take the Calabi-Yau to be large. However, the 
attractor mechanism does not always 
accord us this luxury, since it fixes the near-horizon 
behavior of the moduli.  In the simplest case of \kt,
the size of the horizon is 
large in four dimensional Planck units, provided we choose the charges 
to be large.  However, 
translating to string units, we find that the size is string scale,
because the dilaton is fixed in terms of the charges by virtue of the 
$\CN = 4$ supersymmetry.
To see this, note that the D0-D4 
system is U-dual to a fundamental 
heterotic string with momentum and winding proportional to the charges. 
Since we require that the supergravity solution reproduce the entropy of the 
fundamental string configuration, we are forced to be at the correspondence
point 
\horpol, where the horizon size is string scale. 

For generic Calabi-Yau black holes, $g_s$ is not constrained by the charges, 
so by taking the charges to be large the horizon area
can be made large in both string and Planck units.  
However, in this case there is another effect which may be relevant --
as one increases the charges, the size of some cycles in the Calabi-Yau will
grow.  In general, one might therefore worry that there will be corrections 
due to Kaluza-Klein modes.  Moreover, we would like to 
have a quantitative understanding of higher derivative terms which are 
not of the specific type considered here. 
Although in these cases the resulting equations are much 
more complicated, we hope to report on this in the near future. 

\noindent
{\it Note added:} In \wegotscooped, it was pointed out that the oscillations 
in the numerical solution to \kill, do not die out rapidly enough as $r \to \infty$.
One can check that the asymptotic behaviour of the numerical solution is 
$e^{2 \, g} \sim \cos\(\beta \, r \)/r$,  which resembles a propagating mode.
It was suggested in \wegotscooped\ that an appropriate field redefinition be done to remove this effect  as the propagating mode is a ghost mode, arising as a result 
of the field equations being higher order in derivatives.  A field 
redefinition was suggested based on linearised analysis about flat space.
We note that such a field redefinition will indeed change the asymptotics as desired by removing the offending $ \cos\(\beta \, r \)/r$ piece of the metric function. However, it 
is not clear that at the full non-linear level we will be able to remove the oscillations 
completely. Simple numerical experiments suggest that the oscillations in the bulk 
of the metric (for finite $r$) persist after the certain field redefinitions.  It remains
an open problem to determine the precise field redefinition for the problem at hand
at the non-linear level and check their effect on the physical metric. We hope to return to this issue at a later stage.

\vskip 1cm \centerline{\bf Acknowledgments}
It is a pleasure to thank Mina Aganagic, Raphael Bousso, Bernard de Wit,  
Atish Dabholkar, Sumit Das, Gary Horowitz, Jonathan Hsu, Shamit Kachru, Renata
Kallosh, Ashoke Sen, Al Shapere, Steve Shenker and  Marco Zagermann for discussions. 
VH and MR would in addition like to thank
Atish Dabholkar for an extremely stimulating seminar at Berkeley.  VH
and MR are supported by the  funds from the Berkeley Center for
Theoretical Physics, DOE grant DE-AC03-76SF00098 and
the NSF grant PHY-0098840, and in part by National Science 
Foundation under Grant No. PHY99-07949. AM is supported by the Department of
Energy, under contract DE--AC02--76SF00515.


\listrefs

\end